\documentclass[preprint,5p,twocolumn,10pt]{elsarticle}

\usepackage{graphicx,dcolumn}
\usepackage{bm,amsmath,amsthm,amssymb,amsfonts}
\usepackage{mathtools,cuted}
\usepackage{hyperref}

\journal{Computer Physics Communications}
\biboptions{sort&compress}

\begin{document}

\begin{frontmatter}

\title{A first-principles method to calculate 
       fourth-order elastic constants of solid materials}

\author[a]{Abhiyan Pandit}
\author[a,b]{Angelo Bongiorno\corref{cor}}

\cortext[cor]{Angelo Bongiorno.\\
\textit{angelo.bongiorno@csi.cuny.edu}}

\address[a]{Department of Chemistry, College of Staten Island,
Staten Island, NY 10314, USA}

\address[b]{The Graduate Center of the City
University of New York, New York, NY 10016, USA}

\begin{abstract}
A {\it first-principles} method is presented to 
calculate elastic constants up to the fourth order 
of crystals with the cubic and hexagonal symmetries. 
The method relies on the numerical differentiation 
of the second Piola-Kirchhoff stress tensor and a 
density functional theory approach to compute the 
Cauchy stress tensors for a minimal list of strained 
configurations of a reference state. The number of 
strained configurations required to calculate the 
independent elastic constants of the second, third, and fourth 
order is 24 and 37 for crystals with the cubic and 
hexagonal symmetries, respectively. Here, this method 
is applied to five crystalline materials 
with the cubic symmetry (diamond, silicon, aluminum, 
silver, and gold) and two metals with the 
hexagonal close packing structure (beryllium and magnesium). 
Our results are compared to available experimental 
data and previous computational studies. Calculated linear 
and nonlinear elastic constants are also used, within a 
nonlinear elasticity treatment of a material, to predict 
values of volume and bulk modulus at zero temperature over an 
interval of pressures. To further validate our method, 
these predictions are compared to results obtained from 
explicit density functional theory calculations.
\end{abstract}


\end{frontmatter}

\section{Introduction}

The elastic constants of a material define the 
relationship between stress and applied strain \cite{cl11}. 
The linear coefficients in this relationship correspond 
to the second-order elastic constants (SOECs) \cite{cl11}. 
These coefficients relate to the elastic moduli of a 
material and are important, for example, to quantify 
the linear response to a deformation, and to calculate 
the speed of sound waves. The techniques to measure and calculate 
SOECs are well established, and in fact these coefficients 
are known for a broad class of materials \cite{jca15}. 
Nonlinear elastic constants characterize the anharmonic 
elastic behavior of a material, and they are of both 
fundamental and practical importance as they govern 
how thermoelastic properties change with 
temperature and pressure \cite{cl11,cb20,bb22}. 
The experimental determination of these nonlinear 
elastic coefficients is challenging \cite{cb81,gh84}, 
and computational methods are needed to predict the 
values of these materials parameters \cite{nm85,wl09,vkll16,teks17,kv19,sl20}.
In this work, we present a new method to calculate 
from {\it first principles} elastic constants of 
a material up to the fourth order.

The isothermal third-order elastic constants (TOECs) 
correspond to the first-order anharmonic terms in the 
series expansion of the free energy of the material with 
respect to the Green-Lagrangian strain \cite{cl11}. 
These elastic coefficients characterize the nonlinear 
elastic behavior of a material, and they are related 
to materials properties such as the long-wavelength 
phonon anharmonicities \cite{sss93}, sound 
attenuation \cite{tb64}, the thermodynamic Gr\"{u}neisen 
parameter \cite{pb83,cb20}, thermal expansion and thermal 
conductivity \cite{zm1960,hk81,dcw98}, and the
intrinsic mechanical strength \cite{cl11,ckl12}.
TOECs are typically obtained from acoustoelastic 
experiments \cite{tb64}, wherein sound velocities 
are measured for a material under different stress 
conditions \cite{tb64,kb65,gac78,jd06}. These 
experiments are challenging and subjected to 
error margins \cite{lg11}, and for this reason, 
these coefficients are known for a restricted class of 
materials \cite{teks17,jt68,jh67,na58,hg66}.

The conventional approach to calculate TOECs 
relies on the use of a density functional 
theory (DFT) calculations to construct either energy 
or stress versus strain curves along a number of deformation 
modes (see Ref. \cite{lls21} and references therein). In 
this approach, the whole set of 
linear and nonlinear coefficients 
are then deduced from a nonlinear 
least-square fitting of the energy-strain or stress-strain 
relationships \cite{lls21,wl09,nsg71,nm85,zmg07,cpgc09,hwgd16}.
The application of this method to materials with the 
cubic symmetry is straightforward, as the number of 
independent SOECs and TOECs to be determined is only 
3 and 6, respectively. However, for materials with a lower 
symmetry, this method becomes increasingly cumbersome 
and less attractive, as demonstrated by the very few 
number of applications appeared so far in literature 
(see Ref. \cite{lls21} and references therein).
An alternative approach to calculate TOECs was proposed 
very recently by one of the authors \cite{ccb18}. 
In this method, elastic constants are obtained by 
combining DFT calculations and a finite deformation 
approach \cite{jca15}, where each TOEC is calculated 
independently by second-order numerical differentiation 
of the second Piola-Kirchhoff (PK2) stress tensor \cite{ccb18}.
This method has general applicability, and so far it 
has been applied to both 2D and 3D materials, with 
the cubic, hexagonal, and orthorhombic symmetries \cite{ccb18,bb22}. 
Furthermore, recently this method was used in combination 
with the quasi-harmonic approximation to calculate TOECs 
at finite temperature \cite{bb22}.

Fourth- and higher-order elastic constants 
govern the anharmonic regime of material subjected to 
large deformations \cite{wl09,kv19,sl20,tsgk63,pbg64}.
Knowledge of these higher-order elastic coefficients 
allow to describe and predict mechanical instability 
points of a material \cite{wy93,czljz20}, as well as 
to characterize the nature of elastic phase 
transitions \cite{wl09,sl20}. The experimental determination 
of fourth-order elastic constants (FOECs) is extremely 
challenging, as large uniaxial stresses need to be 
applied in acoustoelastic experiments to obtain 
reliable values of these high-order elastic 
coefficients \cite{cb81,gh84}. For this reason, to the 
best of our knowledge, so far FOECs have been measured 
only for very few materials \cite{cb81,gh84}.
DFT calculations have been employed to calculate FOECs 
\cite{nm85,wl09,vkll16,teks17,kv19,sl20,llz22}. In these 
computational studies, FOECs were obtained by using the 
approach relying on fitting energy-strain or stress-strain 
curves. Although straightforward and in principle general, 
the computational workload and intricacy of this approach 
increase significantly for low-symmetry materials.
Indeed, to the best of our knowledge, to date 
this approach has been used to calculate FOECs of 
materials with only the cubic symmetry 
\cite{nm85,wl09,vkll16,teks17,kv19,sl20}.

In this work, we extend the new method developed to 
calculate TOECs \cite{ccb18} based on finite deformations 
and numerical differentiation of the PK2 stress tensor 
to the calculation of FOECs. The most important advantage 
of the present method over existing approaches 
is that each nonlinear elastic constant is calculated 
independently, by considering up to 8 deformed configurations 
of the reference state. Thanks to this, our method can be 
easily applied to any material, regardless of its symmetry. 
Here we apply the method to calculate SOECs, TOECs, and FOECs 
of five crystalline materials with the cubic symmetry 
(diamond, silicon, aluminum, silver, and gold), and two materials 
with the {\it hcp} structure (magnesium and beryllium).

This manuscript is organized as follows. 
In Sec.\ \ref{method}, we introduce basic notions of 
nonlinear elasticity theory, we provide details 
about the finite difference formulas to calculate 
SOECs, TOECs, and FOECs, and we discuss 
technical aspects of the numerical implementation 
of our methods. In Sec.\ \ref{results}, 
we present results and discuss the application of our 
method to the aforementioned materials.
Conclusions and outlook are provided in 
Sec.\ \ref{ending}.

\section{Methods}\label{method}

\subsection{Notions of nonlinear elasticity theory}

The Green-Lagrangian strain, $\mu_{ij}$, is defined 
as \cite{w67,cl11,ccb18}:
\begin{equation}\label{strain}
\mu_{ij} =  \frac{1}{2} (F_{ki} F_{kj} - \delta_{ij}),
\end{equation}
where subscript indices refer to Cartesian coordinates, 
$\delta_{ij}$ is the Kronecker delta function, 
and $F_{ij}$ are components of the deformation gradient. 
This tensor is defined as:
\begin{equation}\label{deformgradient}
F_{ij} =  \frac{\partial x_i}{\partial X_j} 
\end{equation}
where $x_i$ and $X_i$ are the Cartesian coordinates of 
a material point in the deformed and reference states, 
respectively. 
The Helmholtz free energy density, $A$, can be written as a 
series expansion in terms of the Lagrangian strain 
as follows \cite{w67,wl09,cl11,ccb18,sl20}:
\begin{strip}
\begin{eqnarray}\label{interu}
A &=& \frac{1}{2} \frac{\partial^2 A}
	         {\partial \mu_{ij} \partial\mu_{lm}} \mu_{ij} \mu_{lm}
  + \frac{1}{6} \frac{\partial^3 A} {\partial \mu_{ij} \partial \mu_{lm} 
	  \partial \mu_{pq}} \mu_{ij} \mu_{lm} \mu_{pq} 
  + \frac{1}{24} \frac{\partial^4 A} {\partial \mu_{ij} 
		  \partial \mu_{lm} \partial \mu_{pq}
                  \partial\mu_{rs}} \mu_{ij} \mu_{lm} 
		  \mu_{pq} \mu_{rs} + \cdots \nonumber \\
	&=& \frac{1}{2} C^{(2)}_{ijlm}\mu_{ij} \mu_{lm}
   +\frac{1}{6} C^{(3)}_{ijlmpq} \mu_{ij} \mu_{lm} \mu_{pq}
   +\frac{1}{24} C^{(4)}_{ijlmpqrs}\mu_{ij} 
	\mu_{lm}\mu_{pq}\mu_{rs} + \cdots ,
\end{eqnarray}
\end{strip}
where $C^{(2)}_{ijlm}$, $C^{(3)}_{ijlmpq}$, and $C^{(4)}_{ijlmpqrs}$ 
are the isothermal SOECs, TOECs, and FOECs of the material in the 
reference state, respectively. Given a reference state, the PK2 
stress tensor, $P_{ij}$, can be defined in terms of the 
Helmholtz free energy density, $A$, as:
\begin{equation}\label{pk2}
P_{ij} = \frac{\partial A}{\partial \mu_{ij}},
\end{equation}
whereas the relationship between PK2 and Cauchy 
stress, $\sigma_{ij}$, is \cite{w67,cl11,ccb18,bb22}:
\begin{equation}\label{cpk}
\sigma_{ij} =  \frac{V}{V^\prime} F_{il} P_{lm} F_{jm}, 
\end{equation}
where $V^\prime$ and $V$ are the volumes of the (same) 
material points $\vec{x}$ and $\vec{X}$ in the 
deformed and reference states, respectively.
Eqs. \ref{interu} and \ref{pk2} allow to define 
the relationship between PK2 stress tensor and 
linear and nonlinear elastic constants. Adopting 
the Voigt notation, this relationship takes the 
following form:
\begin{equation}\label{pkecs}
P_{\alpha} = C^{(2)}_{\alpha \beta} \mu_{\beta}
        +\frac{1}{2} C^{(3)}_{\alpha \beta \gamma} 
	 \mu_{\beta} \mu_{\gamma} 
	+ \frac{1}{6} C^{(4)}_{\alpha \beta \gamma \delta} 
	\mu_{\beta} \mu_{\gamma} \mu_{\delta},
\end{equation} 
where Greek indices run from 1 to 6, and are related to 
the Cartesian indices pairs as follows:
1 $\rightarrow$ xx, 2 $\rightarrow$ yy,
3 $\rightarrow$ zz, 4 $\rightarrow$ yz, 5 $\rightarrow$ zx, 
and 6 $\rightarrow$ xy. 
For sake of completeness, 
here below we also express the linear 
and nonlinear elastic constants in terms of the PK2 stress tensor:
\begin{equation}\label{stfecs}
\begin{split}
C^{(2)}_{\alpha \beta} = 
\frac{\partial P_{\alpha}}{\partial \mu_{\beta}} \ &\ \ \
C^{(3)}_{\alpha \beta \gamma} = \frac{\partial^2 P_{\alpha}}
{\partial \mu_{\beta} \mu_{\gamma} } \\ 
C^{(4)}_{\alpha \beta \gamma \delta} &= 
\frac{\partial^3 P_{\alpha}}
{\partial \mu_{\beta} \mu_{\gamma} \mu_{\delta}}.
\end{split}
\end{equation}
The present method relies on the definitions above 
to calculate SOECs, TOECs, and FOECs of a material 
using periodic DFT approach. In this work, temperature 
effects are disregarded and all calculations are 
carried out in static conditions.

\subsection{Finite difference formulas to calculate elastic constants}

To calculate SOECs, we use the following central finite 
difference formula (Eq.\ \ref{stfecs}):
\begin{equation}\label{soecs}
C^{(2)}_{\alpha \beta} = 
    \frac{P_{\alpha}^{\scriptscriptstyle{(+\beta)}}
        - P_{\alpha}^{\scriptscriptstyle{(-\beta)}} }{2 \xi},
\end{equation}
where $\xi$ is a strain parameter, and 
$P_{\alpha}^{\scriptscriptstyle{(\pm\beta)}}$ is the $\alpha$-component 
of the PK2 stress tensor of a deformed configuration 
obtained by applying to the reference state a six-dimensional 
strain vector, $\vec{\mu}$, with component $\beta$ equal to 
$\pm\xi$, and the rest of the components equal to zero.
In case of TOECs, we have two different cases.
TOECs with at least two out of three indices equal to 
each other can be calculated using the following 
second-order central finite difference formula:
\begin{equation}\label{toecs1}
C^{(3)}_{\alpha \beta \beta} = 
\frac{P_{\alpha}^{\scriptscriptstyle{(+\beta)}} + 
P_{\alpha}^{(\scriptscriptstyle{-\beta)}} 
+ 2 P_{\alpha}^{\scriptscriptstyle{(0)}}} {2 \xi^2},
\end{equation}
where $P^{(0)}_{\alpha}$ refers to the $\alpha$-component 
of the PK2 stress tensor of the reference state, which is 
equal to the Cauchy stress tensor.
In case of TOECs whose indices are all different, we 
use the following formula:
\begin{equation}\label{toecs2}
C^{(3)}_{\alpha \beta \gamma} = \frac{
	P_{\alpha}^{\scriptscriptstyle{(+\beta,+\gamma)}} -
	P_{\alpha}^{\scriptscriptstyle{(-\beta,+\gamma)}} -
	P_{\alpha}^{\scriptscriptstyle{(+\beta,-\gamma)}} +
	P_{\alpha}^{\scriptscriptstyle{(-\beta,-\gamma)}} } { 4 \xi^2},
\end{equation} 
where $P_{\alpha}^{\scriptscriptstyle{(\pm\beta,\pm\gamma)}}$ 
is the $\alpha$-component of the PK2 stress tensor of a
deformed configuration obtained by applying to the reference
state a six-dimensional strain vector, $\vec{\mu}$, with 
components $\beta$ and $\gamma$ equal to $\pm\xi$, and the 
rest of the components equal to zero.
In case of FOECs, we have derived the following finite 
difference formulas to calculate the different types of coefficients:
\begin{strip}
\begin{equation}\label{foecs}
\begin{split}
C^{(4)}_{\alpha\beta\beta\beta}&= 
\frac{P^{\scriptscriptstyle{(+2\beta)}}_{\alpha}-
	2 P^{\scriptscriptstyle{(+\beta)}}_{\alpha}+
	2 P^{\scriptscriptstyle{(-\beta)}}_{\alpha}-
	P^{\scriptscriptstyle{(-2\beta)}}_{\alpha}} {2 \xi^3} \\
C^{(4)}_{\alpha\beta\gamma\gamma}&= 
\frac{
P^{\scriptscriptstyle{(+\beta,+2\gamma)}}_{\alpha} -
P^{\scriptscriptstyle{(-\beta,+2\gamma)}}_{\alpha} +
P^{\scriptscriptstyle{(+\beta,-2\gamma)}}_{\alpha} -
P^{\scriptscriptstyle{(-\beta,-2\gamma)}}_{\alpha} -
2 ( P^{\scriptscriptstyle{(+\beta)}}_{\alpha} -
P^{\scriptscriptstyle{(-\beta)}}_{\alpha} ) } {8 \xi^3} \\
C^{(4)}_{\alpha\beta\gamma\delta}&= (
P^{\scriptscriptstyle{(+\beta,+\gamma,+\delta)}}_{\alpha} -
P^{\scriptscriptstyle{(+\beta,+\gamma,-\delta)}}_{\alpha} -
P^{\scriptscriptstyle{(+\beta,-\gamma,+\delta)}}_{\alpha} +
P^{\scriptscriptstyle{(+\beta,-\gamma,-\delta)}}_{\alpha} -
P^{\scriptscriptstyle{(-\beta,+\gamma,+\delta)}}_{\alpha} + \\
	& \qquad \qquad \qquad \qquad \qquad \qquad \qquad \qquad
	+P^{\scriptscriptstyle{(-\beta,+\gamma,-\delta)}}_{\alpha} +
P^{\scriptscriptstyle{(-\beta,-\gamma,+\delta)}}_{\alpha} -
P^{\scriptscriptstyle{(-\beta,-\gamma,-\delta)}}_{\alpha} ) / 8 \xi^3,
\end{split}
\end{equation}
\end{strip}
where $P_{\alpha}^{\scriptscriptstyle{(\pm\beta,\pm\gamma,\pm\delta)}}$
is the $\alpha$-component of the PK2 stress tensor 
of a deformed configuration obtained by applying to the reference state 
a six-dimensional strain vector, 
$\vec{\mu}$, with components $\beta, \gamma$, and $\delta$ equal to $\pm\xi$, 
and the rest of the components equal to zero.

For sake of clarity, we consider the calculation of 
the two nonlinear elastic constants, $C^{(3)}_{123}$ 
and $C^{(4)}_{1255}$, of a material with an arbitrary 
symmetry. Thus, in case of $C^{(3)}_{123}$, 
we consider the following 4 strain vectors:
\begin{eqnarray}
	&& (0, +\xi, +\xi, 0, 0, 0),\  (0, -\xi, +\xi, 0, 0, 0), \nonumber \\
	&& (0, +\xi, -\xi, 0, 0, 0),\  (0, -\xi, -\xi, 0, 0, 0).
\end{eqnarray}
Each strain vector is used to generate a deformed 
configuration of the reference state, and the 
resulting $P_1$ components of the PK2 stress tensors 
are then used in Eq.\ \ref{toecs2} to calculate $C^{(3)}_{123}$.
In case of $C^{(4)}_{1255}$, 
we use the second formula in Eq.\ \ref{foecs}, with 
the component $P_1$ of the PK2 stress tensor 
resulting from the following 6 deformations:
\begin{eqnarray}
&& (0, +\xi, 0, 0, +2\xi, 0) ,\ (0, -\xi, 0, 0, +2\xi, 0), \nonumber \\
&& (0, +\xi, 0, 0, -2\xi, 0) ,\ (0, -\xi, 0, 0, -2\xi, 0), \nonumber \\
&& (0, +\xi, 0, 0,     0, 0) ,\ (0, -\xi, 0, 0,     0, 0).
\end{eqnarray}
These two examples show that, in contrast to conventional 
approaches \cite{wl09,kv19,sl20,czl20,lls21,llz22}, our 
method allows to calculate each nonlinear elastic constant 
independently, regardless of the symmetry of the material.

\subsubsection{SOECs, TOECs, and FOECs of crystals with 
the cubic or hexagonal symmetry}

In this work, we apply our method to materials with the 
cubic and hexagonal symmetry. 
A material belonging to 
the cubic system (point groups: $432$,
$4\bar{3}m$, and $m\bar{3}m$) has 3, 6, and 11 independent SOECs, 
TOECs, and FOECs, respectively \cite{tsgk63,pbg64,cl11,wl09}.
To calculate the 3 independent SOECs, 
we use the following 4 strain vectors:
\begin{eqnarray}\label{soecs_cub}
&& (0, 0, 0, 0, 0, 0), \ (+\xi, 0, 0, 0, 0, 0), \nonumber \\
&& (-\xi, 0, 0, 0, 0, 0), (0, 0, 0, +\xi, 0, 0, 0).
\end{eqnarray}
We highlight that, due to the cubic symmetry, 
$P_{4}^{\scriptscriptstyle{(+4)}}=-P_{4}^{\scriptscriptstyle{(-4)}}$, 
and therefore only one deformation is needed 
to calculate $C_{44}^{(2)}$. To calculate the 6 independent TOECs, 
in addition to the deformations in Eq.\ \ref{soecs_cub}, 
we use the following 4 strain vectors:
\begin{eqnarray}\label{toecs_cub}
	&& (+\xi, +\xi, 0, 0, 0, 0) ,\ (+\xi, -\xi, 0, 0, 0, 0), \nonumber \\
	&& (-\xi, -\xi, 0, 0, 0, 0) ,\ (0, 0, 0, +\xi, +\xi, 0).
\end{eqnarray}
Also in this case, the list above excludes 
strain vectors that lead to redundant 
deformed states of a material with a 
cubic symmetry. The 11 independent FOECs are 
obtained by considering the following 16
additional strain vectors:
\begin{eqnarray}\label{foecs_cub}
&& (+2\xi,  0,  0,  0,  0,  0) ,\ (-2\xi,  0,  0,  0,  0,  0), \nonumber \\
&& (+2\xi, +\xi,  0,  0,  0,  0) ,\ (-2\xi, +\xi,  0,  0,  0,  0), \nonumber \\
&& (+2\xi, -\xi,  0,  0,  0,  0) ,\ (-2\xi, -\xi,  0,  0,  0,  0), \nonumber \\
&& (+\xi,  0,  0, +2\xi,  0,  0) ,\ (-\xi,  0,  0, +2\xi,  0,  0), \nonumber \\
&& (+\xi,  0,  0,  0, +2\xi,  0) ,\ (-\xi,  0,  0,  0, +2\xi,  0), \nonumber \\
&& ( 0,  0,  0, +\xi, +\xi, +\xi) ,\ ( 0,  0,  0, -\xi, +\xi, +\xi), \nonumber \\
&& ( 0,  0,  0, +2\xi,  0,  0) ,\ ( 0,  0,  0, +\xi, +2\xi,  0), \nonumber \\
&& ( 0, +\xi,  0,  0,  0,  0) ,\ ( 0, -\xi,  0,  0,  0,  0).
\end{eqnarray}
In total, to calculate all the independent 
SOECs, TOECs, and FOECs of a material with the 
cubic symmetry (point groups: $432$,
$4\bar{3}m$, and $m\bar{3}m$), our method requires 24
strain vectors (including the null vector for 
the reference state).

A material with the hexagonal symmetry (point groups: $622$, $6mm$, 
$\bar{6}m2$, and $6/mmm$) has 5, 10, and 19 independent SOECs,
TOECs, and FOECs, respectively \cite{cl11,m79}. To calculate 
the 5 independent SOECs, we use the following 6 strain vectors:
\begin{eqnarray}\label{soecs_hex}
&& (0, 0, 0, 0, 0, 0), \ (+\xi, 0, 0, 0, 0, 0) , \nonumber \\
&& (-\xi, 0, 0, 0, 0, 0), \ (0, 0, 0, +\xi, 0, 0), \nonumber \\
&& (0, 0, +\xi, 0, 0, 0), \ (0, 0, -\xi, 0, 0, 0).
\end{eqnarray}
To calculate the 10 independent TOECs, in addition 
to the strain vectors above, we need to account for 
the following 6 strain vectors:
\begin{eqnarray}\label{toecs_hex}
&& (0, +\xi, +\xi, 0, 0, 0), \ (0, -\xi, +\xi, 0, 0, 0) , \nonumber \\
&& (0, +\xi, -\xi, 0, 0, 0), \ (0, -\xi, -\xi, 0, 0, 0), \nonumber \\
&& (0, +\xi, 0, 0, 0, 0), \ (0, -\xi, 0, 0, 0, 0).
\end{eqnarray}
To obtain the 19 independent FOECs, we use the 
following 25 additional strain vectors:
\begin{eqnarray}\label{foecs_hex}
&& (+2\xi, 0, 0, 0, 0, 0), \ (-2\xi, 0, 0, 0, 0, 0) , \nonumber \\
&& (+2\xi, +\xi, 0, 0, 0, 0), \ (+2\xi, -\xi, 0, 0, 0, 0), \nonumber \\
&& (-2\xi, +\xi, 0, 0, 0, 0), \ (-2\xi, -\xi, 0, 0, 0, 0), \nonumber \\
&& (+\xi, 0, +2\xi, 0, 0, 0), \ (-\xi, 0, +2\xi, 0, 0, 0), \nonumber \\
&& (+\xi, 0, -2\xi, 0, 0, 0), \ (-\xi, 0, -2\xi, 0, 0, 0), \nonumber \\
&& (+\xi, 0, 0, +2\xi, 0, 0), \ (-\xi, 0, 0, +2\xi, 0, 0), \nonumber \\
&& (+\xi, 0, 0, 0, +2\xi, 0), \ (-\xi, 0, 0, 0, +2\xi, 0), \nonumber \\
&& (+\xi, 0, 0, 0, 0, +2\xi), \ (-\xi, 0, 0, 0, 0, +2\xi), \nonumber \\
&& (0, +2\xi, +\xi, 0, 0, 0), \ (0, +2\xi, -\xi, 0, 0, 0), \nonumber \\
&& (0, -2\xi, +\xi, 0, 0, 0), \ (0, -2\xi, -\xi, 0, 0, 0), \nonumber \\
&& (0, 0, +2\xi, 0, 0, 0), \ (0, 0, -2\xi, 0, 0, 0), \nonumber \\
&& (0, 0, +\xi, +2\xi, 0, 0), \ (0, 0, -\xi, +2\xi, 0, 0), \nonumber \\
&& (0, 0, 0, +2\xi, 0, 0).
\end{eqnarray}
In total, our method requires 37
strain vectors to calculate all the independent 
SOECs, TOECs, and FOECs of a material belonging 
to the hexagonal crystal system (point groups: $622$, $6mm$,
$\bar{6}m2$, and $6/mmm$).

\subsection{Technical aspects of the method implementation}

Our method to calculate linear and nonlinear 
elastic constants is implemented in codes 
that are part of the software package {\it xPK2x}, 
which is available under the GNU General Public 
License (Version 3) on GitHub \cite{mygit}. 
This software package encompasses three Fortran 
modules, a Bash script, several example 
applications, and relevant documentation \cite{mygit}.
Our method relies on an (external) periodic 
DFT approach to optimize geometries and 
calculate the Cauchy stress tensor. To this 
end, the current version of {\it xPK2x} 
is designed to be compatible with 
the Quantum Espresso software package \cite{qea,qeb}.
For sake of clarity, here below we discuss 
the numerical operations and tasks 
implemented and carried out by the modules 
provided in {\it xPK2x}. 
We refer to the documentation available on 
GitHub \cite{mygit} for additional information 
regarding installation and use of the programs.


The calculation of a set of elastic constants of 
a material requires, as a first step, to select a 
a periodic unit cell to describe
the material in a reference state. The unit 
cell has a volume $V$ and geometry $\bm{V}$:
\begin{equation}\label{ucell}
\bm{V} =
\begin{pmatrix}
a_{1,x} & a_{2,x} & a_{3,x} \\
a_{1,y} & a_{2,y} & a_{3,y} \\
a_{1,z} & a_{2,z} & a_{3,z}
\end{pmatrix},
\end{equation}
where $\vec{a}_1,\vec{a}_2,\vec{a}_3$ are the unit 
cell vectors. We remark that although the choice of 
the reference 
state and corresponding supercell is arbitrary, 
in this work we reports results obtained by considering 
primitive unit cells, and reference states 
yielding a zero static pressure.
Then, given the list of 
elastic constants to be calculated, then 
next operation consists in determining 
the finite difference formulas to be used, 
and therefore list of strain vectors 
required to generate the deformed 
configurations of the reference state. 
Geometry of the reference state and corresponding 
supercell, fractional coordinates of the atoms 
including in it, list of six-dimensional strain 
vectors to generate the deformed configurations, 
and the strain parameter multiplying the strain 
vectors, all these are input parameters for 
the module {\it str2pk} of the software 
package {\it xPK2x} \cite{mygit}. In particular, the 
numerical tasks implemented in the module {\it str2pk} are:

\begin{itemize}

\item Importing the geometry of the reference state 
and (fractional) coordinates of the atoms in the 
supercell (not necessarily a primitive unit cell), 
and reading the list of strain vectors. 
For each strain vector, which we can express in 
both the Voigt and tensorial forms as
\begin{eqnarray}
\vec{\mu} &=& ( \xi_1, \xi_2,  \xi_3, \xi_4, \xi_5, \xi_6 ) \nonumber \\ 
\bm{\mu} &=&
\begin{pmatrix}
\xi_{1}   & \xi_{6}/2 & \xi_{5}/2 \\
\xi_{6}/2 & \xi_{2} &   \xi_{4}/2 \\
\xi_{5}/2 & \xi_{4}/2 & \xi_{3}
\end{pmatrix},
\end{eqnarray}
{\it str2pk} calculates the deformation 
gradient, $\bm{F}$, as follows. 
First, the Cholesky decomposition of the following
3$\times$3 matrix is carried out (see Eq.\ \ref{strain}):
\begin{equation}
2 \bm{\mu} + {\bf I} = \bm{D} \bm{D}^T.
\end{equation}
Then, a single value factorization 
of $\bm{D}$ is carried out, to obtain 
$\bm{D}=\bm{W} \bm{S} \bm{V}^T$,
where $\bm{W}$ and $\bm{V}$ are unitary
matrices, and $\bm{S}$ is the diagonal matrix
of singular values. 
Finally, the rotation-free deformation 
gradient (right stretch tensor) is defined 
as $\bm{F}=\bm{V} \bm{S} \bm{V}^T$ 
($\bm{R}=\bm{W} \bm{V}^T$ is the rotation tensor).

\item Then, the deformation gradient, $\bm{F}$, 
is used to generate the unit cell of the deformed 
configuration by using Eq.\ \ref{deformgradient}. 
In particular, since we consider only 
homogeneous deformations of a material 
described by the use of a periodic unit cell $\bm{V}$,
Eq.\ \ref{deformgradient} assumes the form:
\begin{equation}\label{df_homo}
\bm{F} = \bm{V}^{\prime} \bm{V}^{-1},
\end{equation}
where $\bm{V}^{\prime}$ is the 3$\times$3 matrix 
defining the geometry of the material in the 
deformed state, 
\begin{equation}\label{dcell}
\bm{V}^{\prime} =
\begin{pmatrix}
a^{\prime}_{1,x} & a^{\prime}_{2,x} & a^{\prime}_{3,x} \\
a^{\prime}_{1,y} & a^{\prime}_{2,y} & a^{\prime}_{3,y} \\
a^{\prime}_{1,z} & a^{\prime}_{2,z} & a^{\prime}_{3,z}
\end{pmatrix}.
\end{equation}
Thus, from Eq.\ \ref{df_homo}, the deformed 
configuration is obtained as,
\begin{equation}\label{vddef}
\bm{V}^{\prime} = \bm{F} \bm{V}.
\end{equation}

\item Geometry and dimensions of the unit cells 
describing the deformed configurations, and 
(fractional) coordinates of the atoms in the unit 
cells, are printed out in text files.

\end{itemize}

The next step then consists in using 
a periodic DFT approach \cite{qea,qeb} 
to optimize the geometry of each 
deformed configuration of the reference state, 
and calculate the corresponding Cauchy stress 
tensors, $\bm{\sigma}$. The list of 
Cauchy stress tensors are then supplied to a second 
module, {\it pk2ecs} \cite{mygit}, for the final 
calculation of the desired list of elastic constants. 
In detail, the numerical tasks implemented in this 
module are:

\begin{itemize}

\item For each deformed configuration 
$\bm{V}^{\prime}$, Eq.\ \ref{cpk} is used to 
calculate the PK2 stress tensor from 
the deformation gradient, $\bm{F}$, and 
the calculated Cauchy stress tensor, as follows:
\begin{equation}\label{cpkm1}
\bm{P}  =  \frac{V^{\prime}}{V} \bm{F}^{-1} \bm{\sigma} \bm{F}^{-T},
\end{equation}
where $V^{\prime}$ is the volume of the 
deformed configuration. 

\item This operation is repeated for each 
strain vector, and the corresponding list of 
PK2 stress tensors is finally plugged into 
the finite difference formulas (Eqs. \ref{soecs}-\ref{foecs}) to 
calculate the selected SOECs, TOECs, and FOECs.

\end{itemize}

We remark that the {\it xPK2x} package 
provides the lists of strain vectors 
required to calculate the independent 
SOECs, TOECs, and FOECs of a material with 
the cubic and hexagonal symmetry, and that 
the modules {\it str2pk} and {\it pk2ecs} are 
designed to be user-friendly for these 
classes of materials. However, we also 
remark that the module {\it str2pk} can 
be used to generate any list of strained 
configurations for a reference state of a 
material with an arbitrary symmetry, and that 
the {\it xPK2x} package includes an additional 
module {\it pk2open} that can be adapted and 
extended to the calculation of any elastic 
constant of the second, third, or fourth order. 
Instructions and examples about how to 
combine the modules {\it str2pk} and 
{\it pk2open} can be found on GitHub \cite{mygit}.

\section{Results and discussion}\label{results}

\subsection{Technical details of the DFT calculations}

In this work, we use the ``{\it pw.x}" code of the 
Quantum Espresso package \cite{qea,qeb} to carry out 
DFT calculations, and we use our method to 
calculate the full set of independent SOECs, TOECs, 
and FOECs of diamond, silicon, aluminum, silver, gold, 
beryllium, and magnesium. To describe these materials, 
we use primitive unit cells, and plane-wave energy
cutoffs of 150 and 600 Ry to represent wavefunctions
and electronic charge density, respectively.
In case of Au, we use a local density approximation \cite{pz81}
for the exchange and correlation energy functional, whereas 
the Perdew-Burke-Ernzerhof parametrization \cite{pbe96} of the 
generalized gradient approximation is used for the 
other materials.

To describe the diamond structure of C and Si, we use
ultrasoft psudopotentials ({\it C.pbe-n-rrkjus\_psl.1.0.0.UPF and
Si.pbe-nl-rrkjus\_psl.1.0.0.UPF}) and uniform grids
of 10$\times$10$\times$10 $k$-points to sample the
Brillouin zone. To describe the {\it fcc} 
structure of Ag and Au, we use the ultrasoft 
pseudopotentials ({\it Ag.pbe-spn-rrkjus\_psl.1.0.0.UPF} 
and {\it Au.pz-spn-rrkjus\_psl.1.0.0.UPF}) 
from the Quantum Espresso library: 
\href{https://github.com/dalcorso/pslibrary}
{\it {https://github.com/dalcorso/pslibrary}}, 
whereas in case of {\it fcc} Al, we use a 
norm-conserving psudopotential \cite{tm91} 
generated by using the {\it fhi98PP} software \cite{fs99} 
that was tested and used in a previous study \cite{ccb18}.
To sample the Brillouin zone of the primitive unit cell of
these two metals, we use a uniform grid of 
25$\times$25$\times$25 $k$-points. 
In case of {\it hcp} 
Be and Mg, we use an ultrasoft ({\it Be.pbe-n-rrkjus\_psl.1.0.0.UPF}) 
and a norm-conserving psudopotential \cite{tm91}, 
respectively. The latter psudopotential was generated by 
using the {\it fhi98PP} software \cite{fs99} and 
was tested and used in a previous study \cite{ccb18}.
To sample the Brillouin zones of Be and Mg, 
we use a grid of 20$\times$20$\times$14 $k$-points.
With these technical details, we obtain the 
equilibrium lattice parameters at zero temperature 
reported in Table \ref{table1}. These results are 
in agreement with experimental data.

\begin{table}[ht!]  
\centering
\caption{\label{table1}
Lattice parameters (in \AA) deduced from DFT 
calculations for the crystalline materials 
investigated in this study. Experimental values 
are also reported for comparison.}
\begin{tabular}{ccccc}
\hline
        Crystal & Space group & $a$ &  $c$ & Exp. ($a$/$c$) \\
\hline
        C  & $Fd\bar{3}m$ & 3.57 & - & 3.57 \cite{nm85} \\
        Si & $Fd\bar{3}m$ & 5.47 & - & 5.43 \cite{sbj61,d74} \\
        Al & $Fm\bar{3}m$ & 4.07 & - & 4.03 \cite{sw70} \\
        Ag & $Fm\bar{3}m$ & 4.16 & - & 4.07 \cite{sw70} \\
        Au & $Fm\bar{3}m$ & 4.05 & - & 4.08 \cite{ak04} \\
        Be & $P6_{3}/mmc$ & 2.27 & 3.58 & 2.29/3.58 \cite{mh63} \\
        Mg & $P6_{3}/mmc$ & 3.24 & 5.28 & 3.18/5.15 \cite{emh03} \\
\hline
\end{tabular}
\end{table}

For testing purposes, 
in case of Al, we calculate the nonlinear elastic 
constants for increasing values of the plane-wave 
energy cutoff, as well as for denser grids of 
$k$-points in the Brillouin zone. All DFT calculations 
are carried out by using stringent
convergence criteria: 10$^{-14}$ Ry for selfconsistency 
and 10$^{-6}$ {\it a.u.} for forces.

\subsection{Second- and third-order elastic constants}

\begin{table*}[ht!]
\centering
\caption{\label{table2}
Independent SOECs and TOECs (in GPa) of cubic diamond, 
silicon, aluminum, silver, and gold calculated using 
the method presented in this work. For each material, 
the first row shows our results, the second row 
reports experimental data, and the remaining rows show 
previous results obtained by using the 
conventional approach based on fitting energy-strain 
or stress-strain data points.}
\begin{tabular}{ccccccccccc}
\hline
Crystal & & $C^{(2)}_{11}$ & $C^{(2)}_{12}$ & $C^{(2)}_{44}$ &
   $C^{(3)}_{111}$ & $C^{(3)}_{112}$ & $C^{(3)}_{123}$ & 
   $C^{(3)}_{144}$ & $C^{(3)}_{155}$ & $C^{(3)}_{456}$ \\
\hline
C & This work & 1054 & 124 & 559 &
         -5942 & -1621 & 614 & -200 & -2773 & -1152  \\
 & Exp.\ \cite{teks17} &
          1082 & 125 & 579 &
         -7750 & -2220 & 604 & -1780 & -2800 & -30 \\
 & Ref.\ \cite{ccb18} &
          1037 & 120 & 552 &
         -5876 & -1593 & 618 & -197 & -2739 & -1111 \\
\hline
Si & This work&        153 & 57 & 75 & -751 & -423 & -78 & 16 & -294 & -59 \\
& Exp.\ \cite{jh67} &  166 & 64 & 80 & -795 & -445 & -75 & 15 & -310 & -86 \\
& Ref.\ \cite{ccb18} & 142 & 51 & 72 & -744 & -393 & -59 & 4 & -297 & -59 \\
& Ref.\ \cite{czl20} & 152 & 59 & 78 & -653 & -456 & -96 & 23& -304 & -7 \\
\hline
Al & This work&        103 & 55 & 31 & -1095 & -330 & 44 & -35 & -357 & -14 \\
& Exp.\ \cite{jt68}&   107 & 60 & 28 & -1076 & -315 & 36 & -23 & -340 & -30 \\
& Ref.\ \cite{ccb18} & 108 & 59 & 33 & -1100 & -371 & 104 & 39 & -421 & -22 \\
\hline
Ag &This work&             107 & 79 & 42 & -962 & -566 & -89 & -9 & -444 & 19 \\
& Exp.\ \cite{na58,hg66} & 124 & 94 & 46 & -843 & -529 & 189 & 56 & -637 & 83 \\
& Ref.\ \cite{wl09} &      161 & 119 & 58 & -1012 & -975 & 162 & 80 & -759 & 53 \\
\hline
Au &This work&             207 & 179 & 35 & -1985 & -1177 & -373 & -63 & -749 & 63 \\
& Exp.\ \cite{na58,hg66} & 192 & 163 & 42 & -1729 & -922 & -233 & -13 & -648 & -12 \\
& Ref. \cite{wl09}       & 202 & 174 & 38 & -2023 & -1266 & -263 & -63 & -930 & 54 \\
& Ref.\ \cite{llz22}     & 151 & 126 & 28 & -1438 & -875 & -550 & -66 & -469 & 16 \\
\hline
\end{tabular}
\end{table*}

\begin{table*}[ht!]
\centering
\caption{\label{table3}
Independent SOECs and TOECs (in GPa) of {\it hcp} 
beryllium and magnesium.
For each crystal, the first row shows our results, 
the second row reports experimental data, and the remaining 
rows show previous results obtained by using the 
conventional approach. }
\begin{tabular}{cccccccccccccccc}
\hline
 & $C^{(2)}_{11}$ & $C^{(2)}_{12}$ & $C^{(2)}_{13}$ &
   $C^{(2)}_{33}$ & $C^{(2)}_{44}$ & $C^{(3)}_{111}$ &
   $C^{(3)}_{112}$ & $C^{(3)}_{113}$ & $C^{(3)}_{123}$ &
   $C^{(3)}_{133}$ & $C^{(3)}_{144}$ &  $C^{(3)}_{155}$ &
   $C^{(3)}_{222}$ & $C^{(3)}_{333}$ &  $C^{(3)}_{344}$ \\
\hline
	\multicolumn{16}{c}{Be}\\
\hline
This work       & 275 & 40 & 30 & 309 & 141 & -3160&211&33&-170&52&-139&-344&-2414&-3826&-948\\
Exp. \cite{ml04}& 294 & 27 & 14 & 357 & 162 & --&--&--&--&--&--&--&--&--&--\\
Ref. \cite{jwi17}&333&16&5&392&171&-5093&1187&707&-87&-838&-435 &-475&-2845&-2048&-489\\
\hline
	\multicolumn{16}{c}{Mg}\\
\hline
This work &      54& 23 & 17 & 58 & 15 &-702& -31&-1&-43&-101&-21&-72&-546&-619&-155 \\
Exp. \cite{n91}& 59& 26 & -- & 62 & 16 &-663&-178&30&-76&-86 &-30&-58&-864&-726&-193\\
Ref.\ \cite{ccb18}&58&24&19&62&16&-602&-190&4&-55&-107&-60&-50&-762&-657&-163\\
Ref.\ \cite{lls21}&68&28&20&70&18&-784&-241&97&-46&-116&-52&-29&-1081&-554&-154\\
\hline
\end{tabular}
\end{table*}

\begin{figure}[!ht]
\centering
\includegraphics[width=\columnwidth]{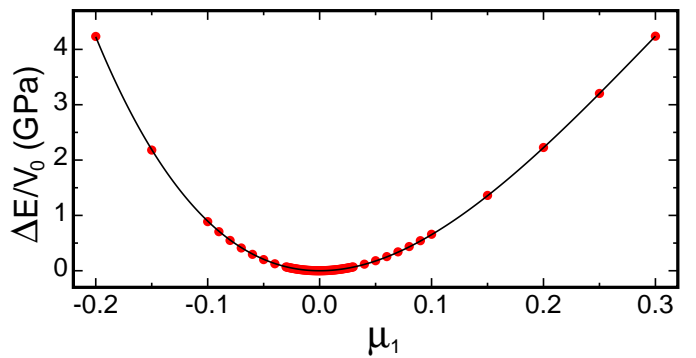}
\caption{\label{fig1} Energy density of cubic Si 
relative to that one of the reference state versus uniaxial 
Lagrangian strain. The solid black line shows the 
fifth-order polynomial function fitting the data (red discs) 
calculated from DFT. $V_0$ is the volume of the reference 
state, whereas $\mu_1$ is the first component of the strain 
tensor in Voigt notation; the remaining components are zero.}
\end{figure}

The independent SOECs and TOECs of crystals 
with the cubic and {\it hcp} structures 
calculated using our method are listed in 
Tables \ref{table2} and \ref{table3}, respectively.
These tables report also available experimental data 
and previous values calculated by using 
the conventional approach relying on fitting 
energy-strain or stress-strain
curves \cite{nm85,dg95,wl09,vkll16,teks17,kv19,sl20}.
We remark that our results are in overall 
good agreement with both experimental data and 
previous computational studies. It is to be 
noted that measurements of TOECs are typically 
carried out at finite temperature, and that sample 
microstructure and defects are known to affect 
to some extent the experimental data \cite{wl09,teks17}.
We attribute to these two factors the origin 
of the small differences between our results and 
the experimental data. As for the differences 
between our results and those of previous 
computational studies, we argue that these 
stem mainly from the following two reasons. 
One, the technical aspects 
of the DFT calculations, namely
plane-wave energy cutoffs, pseudopotentials, 
convergence thresholds, and the exchange and 
correlation energy functional. Two, 
the details of the fitting procedure used 
to deduce the full set of independent linear 
and nonlinear elastic constants \cite{czl20}.
To corroborate this argument, and at the same 
time, to demonstrate the validity of our 
method and results, we adopt the conventional 
approach based on fitting an energy-strain curve 
to calculate the elastic constants $C^{(2)}_{11}$, 
$C^{(3)}_{111}$, $C^{(4)}_{1111}$, and $C^{(5)}_{11111}$ 
of Si (Fig.\ \ref{fig1}). To this end, we use 
a fifth-order polynomial function to fit the 
energy versus strain data points computed 
from DFT for a set of deformed configurations of Si 
obtained by applying a uniaxial strain along the $x$ 
direction (Fig.\ \ref{fig1}). The fitting procedure 
yields the following values: 
$C^{(2)}_{11}$ = 153 GPa , $C^{(3)}_{111}$ = -730 GPa ,
$C^{(4)}_{1111}$ = 2555 GPa , $C^{(5)}_{11111}$ = -10493 GPa.
These values are in excellent agreement with the 
elastic constants computed by using the present 
method reported in Tables \ref{table2} and \ref{table4}.

\subsection{Fourth-order elastic constants}

\begin{figure}[!ht]
\centering
\includegraphics[width=\columnwidth]{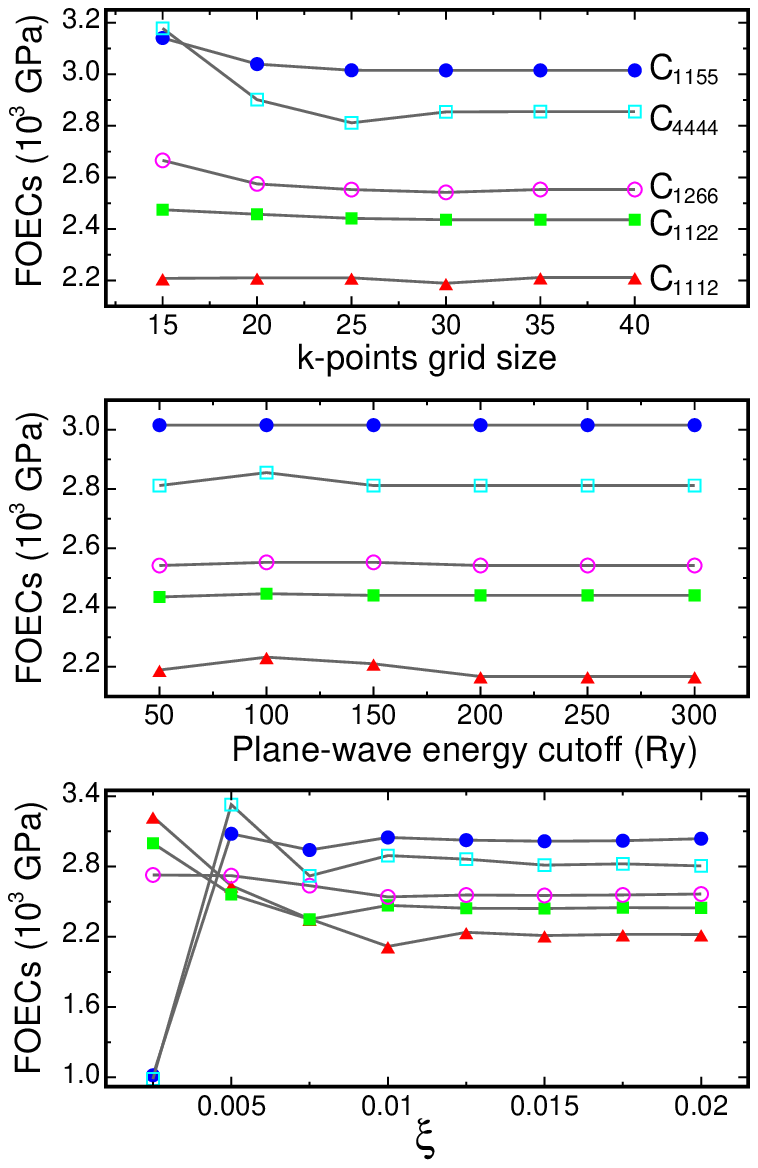}
\caption{\label{fig2} Values of $C^{(4)}_{1112}$ 
(red filled triangles), $C^{(4)}_{1122}$ (green filled 
squares), $C^{(4)}_{1155}$ (blue filled circles),
$C^{(4)}_{1266}$ (magenta open circles), and $C^{(4)}_{4444}$ 
(cyan open squares) of {\it} fcc Al calculated from DFT by 
considering (top panel) uniform 
grids of $k$-points of increasing density 
(and a fixed plane-wave energy cutoff of 150 Ry and 
a strain parameter of 0.015), 
(middle panel) increasing values of the 
plane-wave energy cutoff used to represent 
wavefunctions (and a fixed 25$\times$25$\times$25 
grid of $k$-points and a strain parameter of 0.015), 
and (bottom panel) increasing 
values of the strain parameter. These last calculations 
are carried out using a plane-wave energy cutoff of 150 Ry 
and a 25$\times$25$\times$25 grid of $k$-points.}
\end{figure}

To assess the accuracy of our results, we carry out convergence tests 
for the selected FOECs of {\it fcc} Al as a function of 
the strain parameter ($\xi$), and also by considering DFT calculations 
of increasing precision (Fig.\ \ref{fig2}). The results of 
these calculations show that FOECs (and TOECs) converge 
rapidly for increasing values of both the $k$-points grid density and 
plane-wave energy cutoff. Also, these calculations show 
that FOECs are sensitive to the value of the strain 
parameter used to generate the deformed configurations of 
a reference state. In particular, Fig.\ \ref{fig2} shows 
that while several FOECs fluctuate 
significantly for strain parameters smaller than 0.0075, 
all the independent FOECs converge and plateau 
for strain parameters larger than 0.01.

Table \ref{table4} reports calculated values of FOECs 
of diamond, Si, Al, Ag, and Au. To the best of our knowledge, 
experimental data for these coefficients are missing 
from literature. Values of FOECs obtained using the 
present method are in reasonable agreement with previous 
results obtained by employing the conventional approach.
We remark that our method yields results 
in excellent agreement with FOECs obtained by fitting 
energy-strain curves. In fact, as discussed above, these 
two methods yield values of $C^{(4)}_{1111}$ for Si 
equal to 2586 and 2555 GPa, respectively. Therefore, once 
again we are inclined to attribute the differences 
between our results and previous calculations \cite{wl09, teks17, czl20, llz22} 
to both different technicalities of the DFT calculations 
and details of the fitting procedure.

It is interesting to notice that Hiki {\it et al.} \cite{hg66,htg67} 
suggested that ``{\it the contribution from the closed-shell 
repulsive interaction between nearest-neighbor ions becomes 
predominant for determining 
the higher order elastic constants for materials with markedly 
overlapped closed shells}'', and therefore that 
FOECs of metals such as Ag and Au should obey the following 
approximate relationships:
\begin{equation}\label{foecs-relation}
\begin{split}
C^{(4)}_{1111} &= 2C^{(4)}_{1112} = 2C^{(4)}_{1122} = 2C^{(4)}_{1155} 
= 2C^{(4)}_{1266} = 2C^{(4)}_{4444} \\ 
C^{(4)}_{1123} &= C^{(4)}_{1144} = C^{(4)}_{1255} = C^{(4)}_{1456} = C^{(4)}_{4455} = 0.
\end{split}
\end{equation}
Using our values for Ag in Table \ref{table4}, we find 
$C^{(4)}_{1111}$/$C^{(4)}_{1112}$=1.9, 
$C^{(4)}_{1111}$/$C^{(4)}_{1122}$=2.0, 
$C^{(4)}_{1111}$/$C^{(4)}_{1155}$=2.2,
$C^{(4)}_{1111}$/$C^{(4)}_{1266}$=2.2,
and $C^{(4)}_{1111}$/$C^{(4)}_{4444}$=2.3, i.e. 
all values close to 2.0, whereas the remaining FOECs 
are much smaller than $C^{(4)}_{1111}$ and thus negligible.
This result not only corroborates the argument put forward by 
Hiki {\it et al.} \cite{hg66,htg67}, but it further 
validates the correctness of our method.

Existing methods based on 
fitting energy-strain or stress-strain curves become 
cumbersome and difficult to apply in case of materials with 
a symmetry lower than the cubic. In contrast, our method is 
easily applicable to materials of any symmetry, and 
the computational workload increases only moderately as 
the symmetry of the material decreases. Here, to demonstrate 
the potential of the present method, we calculate 
the independent FOECs of {\it hcp} Be and Mg. The 
results of these calculations are shown in Table \ref{table5}.
To the best of our knowledge, FOECs of 
these two materials have so far neither been measured nor 
calculated. 

\begin{table*}[ht!]
\centering
\caption{\label{table4}
Independent FOECs (in GPa) of cubic diamond, silicon, 
aluminum, silver, and gold obtained by using the 
present method. Our results are compared to values 
calculated by employing the conventional approach 
relying on fitting energy-strain curves. }
\begin{tabular}{ccccccccccccc}
\hline
 & & $C^{(4)}_{1111}$ & $C^{(4)}_{1112}$ & $C^{(4)}_{1122}$ &
   $C^{(4)}_{1123}$ & $C^{(4)}_{1144}$ & $C^{(4)}_{1155}$ &
   $C^{(4)}_{1255}$ & $C^{(4)}_{1266}$ & $C^{(4)}_{1456}$ &
   $C^{(4)}_{4444}$ & $C^{(4)}_{4455}$\\
\hline
C & This work & 36057 & 9864 & 6768 & -519 & -1747 & 12628 & 284 & 9662 & 1236 & 12926 & 1169 \\
  & Ref.\ \cite{teks17} & 26687 & 9459 & 6074 & -425 & -1385 & 10741 & -264 & 8192 & 487 & 11328 & 528 \\
\hline 
Si & This work & 2586 & 2112 & 1885 & 576 & -671 & 833 & -422 & 742 & -46 & 1268 & -2 \\
   & Ref.\ \cite{czl20} & 613  & 2401 & 1275 & 1053 & 5071 & 4050 & -2728 & -514 & 66 & -2553 & -577 \\
\hline
Al & This work & 10102 & 2210 & 2441 & -609 & -68 & 3016 & 159 & 2553 & 224 & 2812 & 180 \\
 & Ref.\ \cite{wl09} & 9916 & 2656 & 3708 & -1000 & -578 & 3554 & -91 & 4309 & 148 & 3329 & 127 \\
\hline
Ag & This work & 8346 & 4429 & 4204 & 333 & 99 & 3735 & 21 & 3813 & -39 & 3638 & -86 \\
 & Ref.\ \cite{wl09} & 13694 & 7115 & 6652 & -387 & -154 & 5295 & 3 & 6718 & -196 & 5416 & -75 \\
\hline
Au & This work & 17113 & 8114 & 8814 & 874 & 860 & 7462 & -634 & 7372 & -257 & 8258 & -61 \\
& Ref. \cite{wl09} & 17951 & 8729 & 9033 & 416 & 691 & 7774 & -752 & 9402 & -170 & 8352 & 15 \\
& Ref. \cite{llz22} & 10094 & 8280 & 8402 & 1507 & 235 & 5549 & -1534 & 8252 & 2 & 3640 & -5763 \\
\hline
\end{tabular}
\end{table*}

\begin{table*}[ht!]
\centering
\caption{\label{table5}
Independent FOECs (in GPa) of {\it hcp} beryllium and magnesium 
calculated by using the present method.}
\begin{tabular}{ccccccccccc}
\hline
 & $C^{(4)}_{1111}$ & $C^{(4)}_{1112}$ & $C^{(4)}_{1113}$ &
   $C^{(4)}_{1122}$ & $C^{(4)}_{1133}$ & $C^{(4)}_{1123}$ &
   $C^{(4)}_{1144}$ & $C^{(4)}_{1155}$ & $C^{(4)}_{1166}$ &
   $C^{(4)}_{1223}$  \\
\hline
Be & 32466 & -3 & 358 & -3529 & -3721 & 881 &
        -1902 & -1342 & -2880 & 1770 \\

Mg & 8638 & -79 & -243 & 119 & -57 & -47 & 
        -69 & -40 & -188 & 266 \\
\hline
& $C^{(4)}_{1233}$ &  $C^{(4)}_{1244}$ &
  $C^{(4)}_{1255}$ & $C^{(4)}_{1333}$ &  $C^{(4)}_{1344}$ &
  $C^{(4)}_{1355}$ & $C^{(4)}_{3333}$ &  $C^{(4)}_{3344}$ &
  $C^{(4)}_{4444}$ & \\
\hline
Be & -2113 & 3838 & 18 & 9934 & 229 &
          1629 & 9986 & 8380 & -5202 & \\

Mg & 347 & 353 & -30 & 828 & 392 & 240 &
	5684 & 1402 & -1073 & \\
\hline
\end{tabular}
\end{table*}

\subsection{Potential application of our method}

Fourth- and higher-order elastic constants
describe the elastic response of a material 
subjected to large deformations \cite{wl09,kv19,sl20,tsgk63,pbg64}.
Knowledge of these higher-order elastic 
coefficients can be thus used to predict, 
within the context of a nonlinear elasticity 
theory treatment, both the strain response and 
SOECs of a material subjected to an external 
pressure (or stress). In this section, we show 
that indeed SOECs, TOECs, and most importantly, 
FOECs, can be used for this purpose, and that 
FOECs expand the predictive power of the 
numerical framework relying on nonlinear elasticity 
theory to larger intervals of strain and pressures.
Here we show the results obtained for {\it fcc} Si 
and {\it hcp} Mg. 

We use both DFT calculations and nonlinear 
elasticity theory to calculate the volume, $V(p)$, and                    
bulk modulus, $B_0(p)$, of Si and Mg at zero 
temperature over a finite interval of pressures. 
In detail, we use variable-cell optimization 
calculations \cite{qea,qeb} and the finite difference 
formulas in Eq.\ \ref{soecs} to calculate from DFT, 
first the volume, and then the SOECs of Si and Mg 
at a pressure $p$. To calculate $B_0(p)$ 
of {\it fcc} Si and {\it hcp} Mg, we use the 
formulas \cite{w67,jzn11,ycq15}
\begin{equation}
B_{0}(p) = \frac{C^{(2)}_{11}(p) + 2C^{(2)}_{12}(p) + p}{3}
\end{equation}
and
\begin{equation}
B_{0} = \frac{2 (C^{(2)}_{11}(p) + C^{(2)}_{12}(p)) + 
C^{(2)}_{33}(p) + 4 C^{(2)}_{13}(p) + 3 p }{9},
\end{equation}
respectively.
We also calculate the same quantities, $V(p)$ and $B_0(p)$, 
within the context of nonlinear elasticity theory by employing 
elastic coefficients calculated with the present 
method. In particular, we use the values of SOECs, TOECs, 
and FOECs for Si and Mg reported in 
Tables \ref{table2}-\ref{table5}. We underline that 
these coefficients are obtained by considering 
a reference state yielding a zero 
static pressure at zero temperature.
Then, we use a self-consistent variational approach to 
solve Eqs.\ \ref{cpk} and \ref{pkecs} 
and determine the strain required to deform the 
reference state and obtain a configuration for the 
material, $V(p)$, yielding a pressure $p$ \cite{bb22}. 
After determining the geometry of the 
material at $p$, we proceed to calculate the SOECs 
and therefore the bulk modulus $B_0(p)$ 
using the same approach relying on the 
finite difference formulas in Eq.\ \ref{soecs}.
However, in this case, the Cauchy and hence 
PK2 stress tensor resulting from a deformation of 
the state $V(p)$ is not calculated  explicitly 
from DFT, but instead it is again derived from 
Eqs.\ \ref{cpk} and \ref{pkecs} as outlined in 
the following diagram:
\begin{equation}\label{meth}
\begin{split}
V(p) \xrightarrow{\tilde{\bm{\mu}}} &
\tilde{\bm{F}},\tilde{V}
\xrightarrow{V(p)} \bm{\mu},\bm{F} 
\xrightarrow{\bm{\mu}} \bm{P}(\bm{\mu}) 
\xrightarrow{\bm{F}} \ldots \\
& \ldots \xrightarrow{\bm{F}} \bm{\sigma}(\bm{\mu}) =
\tilde{\bm{\sigma}}(\tilde{\bm{\mu}})
\xrightarrow{\tilde{\bm{F}}} \tilde{\bm{P}}(\bm{\tilde{\mu}}),
\end{split}
\end{equation}
where $\tilde{\bm{\mu}}$ and $\tilde{\bm{F}}$ are the
Lagrangian strain and corresponding deformation gradient
mapping $V(p)$ to one of its deformed states,
$\tilde{V}$, whereas $\bm{\mu}$ and $\bm{F}$ are
the strain and deformation gradient mapping $V(p)$
to $\tilde{V}$. Thanks to this last correspondence,
Eq.\ \ref{pkecs} can be used to
extrapolate the value of the PK2 stress tensor
in $\tilde{V}$ resulting from the deformation
of $V(p)$, whereas Eq.\ \ref{cpk} can be used
to, first, calculate the Cauchy stress,
$\bm{\sigma}(\bm{\mu})=\tilde{\bm{\sigma}}(\tilde{\bm{\mu}})$,
and then the PK2 stress tensor resulting from the
deformation of $V(p)$, which is needed to
calculate its SOECs.

\begin{figure}[!ht]
\centering
\includegraphics[width=\columnwidth]{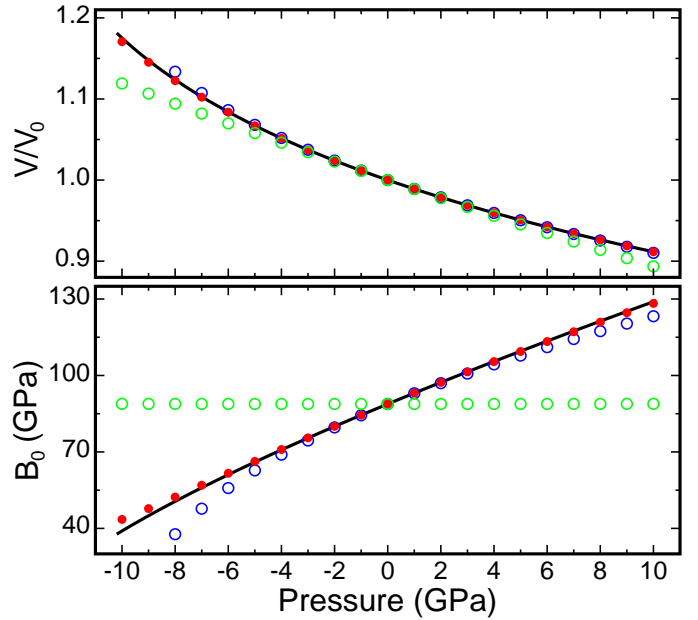}
\caption{\label{fig3} Top panel, unit-cell volume relative to 
that one at zero pressure and, bottom panel, bulk 
modulus of cubic Si versus pressure. Black solid line 
shows results obtained from DFT calculatios, whereas 
discs and circles show results obtained from nonlinear 
elasticity theory: green circles, blue circles, and red 
discs show results obtained by considering only 
SOECs, SOECS and TOECs, and all the elastic constants 
up to FOECs, respectively.}
\end{figure}

\begin{figure}[!ht]
\centering
\includegraphics[width=\columnwidth]{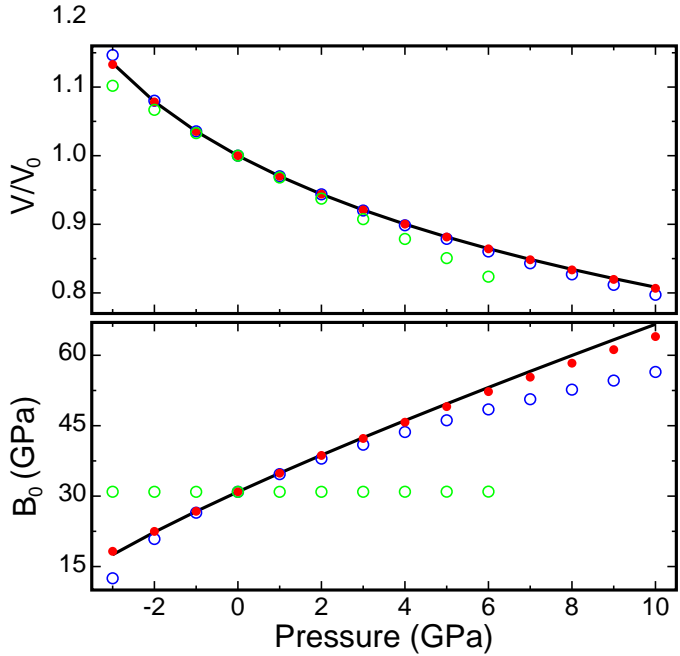}
\caption{\label{fig4} Same as Fig.\ \ref{fig3} 
for {\it hcp} Mg}
\end{figure}

The results of these two sets of calculations 
are compared in Figs.\ \ref{fig3} and \ref{fig4} for 
Si and Mg, respectively.
These comparisons show, as expected, that 
the formalism relying on nonlinear elasticity theory 
yields results that agree with those obtained from DFT 
over larger intervals of pressure for increasing the 
order of the truncation in Eq.\ \ref{pkecs}, i.e. 
considering the higher-order elastic constants. 
In particular, while in case of the equation of 
state $V(p)$, a good agreement is already reached 
by considering only SOECs and TOECs, in case of 
$B_0(p)$, the inclusion of FOECs in Eq.\ \ref{pkecs} 
is necessary to achieve an excellent agreement 
over the full intervals of pressures.
 
\section{Conclusion}\label{ending}

We presented a method to
calculate second-, third-, and fourth-order elastic
constants of crystals with the cubic and hexagonal
symmetry. This {\it first-principles} 
method relies on the numerical
differentiation of the second Piola-Kirchhoff stress
tensor and a minimal list of strained configurations 
of a reference state for a material.
In particular, the number of configurations required 
to calculate the independent elastic constants up to 
the fourth order is 24 and 37 for a crystal
with the cubic and hexagonal symmetry, respectively.
Although here we have shown applications to materials 
with the cubic and hexagonal symmetry, our method 
has general applicability as, regardless of 
symmetry, each elastic constant of any order can be
calculated independently by carrying out several DFT 
calculations. This important aspect is what differentiates 
our method from conventional approaches based on 
fitting energy-strain or stress-strain curves.
 
To validate our method, here we calculated the elastic 
constants up to the fourth order of five and two 
materials with the {\it fcc} and {\it hcp} structures, 
respectively. Comparisons of our results with available 
experimental data and previous calculations show that 
our method is reliable and accurate.
We have also used a formalism based on nonlinear 
elasticity theory to predict the equation of state 
and elastic properties of a material 
over finite intervals of pressure. This formalism 
requires as input parameters linear and nonlinear 
elastic constants of a material in a reference state, 
and its predictive power improves as higher-order 
elastic constants are accounted for.
Our method has the potential to be extended to the 
calculation of elastic constants of the fifth or 
higher order of a material with an arbitrary symmetry.
Therefore, the present method has the potential 
to enhance the capabilities of the aforementioned
formalism based on nonlinear elasticity theory 
to predict, for example, thermoelastic 
behaviors \cite{bb22}, the occurrence of 
solid phase transitions \cite{czl20}, and values 
of ideal yield strengths \cite{czl20}.

\section{Acknowledgements}

This work is supported by the National Science
Foundation (NSF), Award No. DMR-2036176.
We acknowledge the support of the CUNY High Performance
Computing Center, the PSC-CUNY grants 62651-0050 and
63913-0051.


\end{document}